\documentclass{article}

\usepackage{PRIMEarxiv}

\usepackage[utf8]{inputenc} 
\usepackage[T1]{fontenc}    
\usepackage{hyperref}       
\usepackage{url}            
\usepackage{booktabs}       
\usepackage{amsfonts}       
\usepackage{nicefrac}       
\usepackage{microtype}      
\usepackage{lipsum}
\usepackage{fancyhdr}       
\usepackage{graphicx}       
\graphicspath{{media/}}     

\title{Reviewer Preferences and Gender Disparities  \newline in Aesthetic Judgments
\thanks{\textit{\underline{Citation}}: 
\textbf{Lassen, I.M.S, Bizzoni, Y., Peura, ., Thomsen, R., \& Nielbo, K.L. . Reviewer Preferences and Gender Disparities in Aesthetic Judgments. 	arXiv:xxxx.xxxxx [cs.CY]}}
}

\author{
  Ida Marie S. Lassen, Yuri Bizzoni, Telma Peura\\
  Center for Humanities Computing Aarhus\\
  Jens Chr. Skous Vej 4, Building 1483\\
  Aarhus University\\
  DK-8000 Aarhus C\\
  \texttt{idamarie@cas.au.dk}, \texttt{yuri.bizzoni@cc.au.dk}, \texttt{tpeura@cc.au.dk} \\
   \And
  Mads  Rosendahl Thomsen\\
  School of Communication and Culture - Comparative Literature\\
  Aarhus University\\
  Langelandsgade 139, building 1580\\
  \texttt{madsrt@cc.au.dk} \\
  \AND
  Kristoffer Nielbo \\
  Center for Humanities Computing Aarhus \& Interacting Minds Centre\\
  Aarhus University\\
  Jens Chr. Skous Vej 4, Building 1483, 3rd floor, DK-8000 Aarhus C\\
  \texttt{kln@cas.au.dk} \\
}

\begin{document}
\maketitle

\begin{abstract}

Aesthetic preferences are considered highly subjective resulting in inherently noisy judgements of aesthetic objects, yet certain aspects of aesthetic judgement display convergent trends over time.  This paper present a study that uses literary reviews as a proxy for aesthetic judgement in order to identify systematic components that can be attributed to bias. Specifically we find that judgement of literary quality in newspapers displays a gender bias in preference of male writers. Male reviewers have a same gender preference while female reviewer show an opposite gender preference. While alternative accounts exist of this apparent gender disparity, we argue that it reflects a cultural gender antagonism.

\end{abstract}


\section{Introduction}

Aesthetic judgements are notoriously complex and subject to considerable variation, because aesthetic objects are complex (ex. literature is a complex linguistic phenomenon that conveys information indirectly), aesthetic preferences are subjective (ex. readers have different aesthetic preferences), and there is a general lack of robust scientific measuring instruments (ex. there is no definitive metric to measure aesthetics or aesthetic judgments). Literary quality, for instance, can be considered one of the most subjective fields of evaluation, and variation mostly attributable to noise introduced by individual preferences. Yet the perception of literary quality from large amounts of readers over time does show convergent trends: communities tend to establish and update canons\cite{guillory1993cultural}; specific texts and narratives manage to remain popular\cite{stephens2013retelling} despite the changing of fashions and political phases and certain author names become eponymous of literary quality in different countries and throughout the social spectrum\cite{bloom2014western}. Some facets of literary quality can be explained in terms of the literary content (ex. predictability of content, coherence of the narrative), while others depend on socio-cultural priors that introduce systematic variation in aesthetic judgements. It is the latter that are the object of this study, specifically the possible effects of gender on assessment of literary quality as an example of how aesthetic judgements can be biased by contextual factors.

There are two important caveats to consider. First, we are not claiming that variability in aesthetic judgement is undesirable, on the contrary, it facilitates expressive variation and counters aesthetic standardization as has been the norm under some authoritarian regimes \cite{buch2016composing, frajese2006nascita, herrero2012censorship, kwon1991literature}. We are only interested in the systematic components of aesthetic judgement that can be attributed to bias, specifically gender bias, and approach this problem from the perspective of fairness challenges in classification of real-world data \cite{miconi_impossibility_2017}. Second, it is not our intention to `point fingers' or addressing specific individual (ex. specific reviewers) or institutional levels of biases (ex. specific outlets). Fairness challenges first and foremost concern a systemic level of biases, that is, macro-relations that are systematic and disadvantages groups of people based on their identity (gender, race, class, sexual orientation), while at the same time advantaging members of a dominant group. While at the individual level a bias effect may seem small or trivial, it is important to emphasize that systems of bias can result in rampant injustice \cite{kahneman_noise_2021}.

The problem of literary quality's subjective status becomes even more intriguing when we turn to the challenge of its computational assessment. Most studies assume the possibility of one one-dimensional ground truth by modelling literary quality as a single rating  or class associated with a text \cite{ferrer2013canonical, wang2019success, walsh2021goodreads}. These ground truths are retrieved from various sources: literary critics, book sale numbers, bestseller lists, or crowd-sourced reader opinions. Such approaches have several limitations: Relying only on experts' judgment (ex. awards, prestigious reviews) biases the model to reflect their preferences, but striving for representativity by crowd-sourcing opinions ends up ignoring important differences in the readers' population. To properly understand the scientific value of these ground truths and develop standardized measures of quality, it is necessary to model possible sources of bias. 

Recent studies have analyzed the impact of the gender of authors as well as of reviewers in literary reviews. \cite{touileb-etal-2020-gender} investigates differences in sentiments in Norwegian book reviews and how literary reviewers are describing authors of same and opposite gender. Their findings show differences in how female and male book authors are positively or negatively described and that the gender of the critics influences this difference. 
In line with the findings in \cite{thelwall_2019} for Goodreads reviews, the authors point out, that male critics deem crime novels written by female authors and sentimental romance novel by male authors as negative and suggest that this indicate that book reviews contain the social hierarchies tending to ascribe emotional traits to women. In the Goodreads reviews, differences are both found in preferences of types of books as well as within genres, meaning that when reviewers of both genders read and review books of the same genre, differences in grading are found between male and female reviewers. In addition, the results show that within the majority of genres, readers prefer books written by an author of their own gender.

In the greater context of circulation and reception of books, \cite{squires2020review} and \cite{dane2020gender} address the role of both review and reviewer in the broader Anglophone literary field. The former point to an imbalance found in the British and Australian review scenes: Most book reviewers are men, and books reviewed are often written by men, resulting in books written by female authors are being treated like a niche.  The latter offers a historical account of the gendered structure of the literary field and maps out how authors build their reputations and accumulate prestige in contemporary book publishing. By taking the historical perspectives of the literary field into account, it might not be surprising that gender disparities still exist. As the reception and judgment of books exist within a greater societal context where structural oppression occurs, signs of systemic inequality call for further investigation. 

In other areas, studies have examined the role of gender in assessment situations. \cite{macnell2015s} shows how students' ratings of instructors are biased towards a more positive assessment of male instructors compared to female instructors. By conducting their study in an online learning situation, the authors were able to disguise the gender identity of the instructors. The found bias was not dependent on the actual gender of the instructor, but on the perceived gender of the instructor. That allows for a conclusion that points out that the gender bias is not a result of gendered behavior of the instructors, but actual bias in the students, suggesting that a female instructor would have to work harder than a male to receive comparable ratings. In an academic context, a study from 2020 \cite{johnson2020dual} shows how female applicants were less likely than male applicants to receive access to resources (in terms of telescope time) when the review process was single-, rather than dual-anonymized. In particular, the findings indicate that male reviewers rated female applicants significantly worse than they rated male applicants before dual--anonymization was adopted, and after applying dual--anonymization, the gender bias was reduced. Similar results are shown in the hiring process of orchestra musicians \cite{goldin2000orchestrating} and several studies have shed light upon the effect of gender in hiring processes \cite{booth2010employers, cole2004interaction}.

Evidence of gender bias across domains, may indicate that similar structural dynamics are at play, and hence, not a unique gender bias evolving in the field of literature.

\section{Methods}\label{sec:methods}

\subsection{Data}

\begin{figure}

\begin{floatrow}
\ffigbox{%
  \includegraphics[width=8cm]{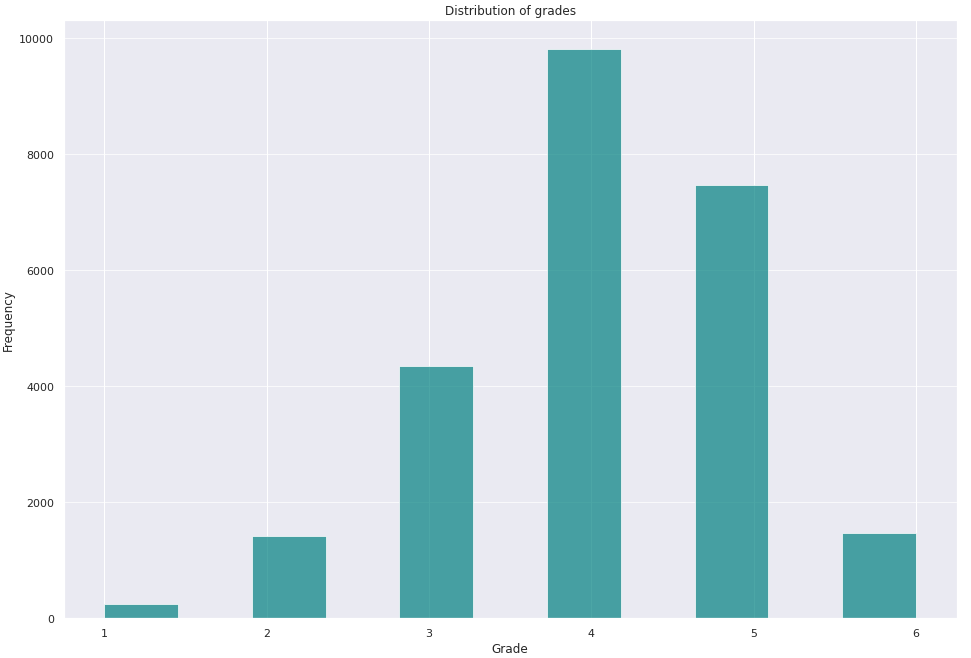} %
}{%
  \caption{Histogram over grades given in Danish newspapers, after grades are transformed into shared 6-point likert scale}%
  \label{fig:dist}

}
\capbtabbox{%
\newcommand*{\MyIndent}{\hspace*{0.5cm}}
\begin{tabular}{||c  c |} 
 \hline
 \MyIndent \textbf{Dataset overview} &    \\ [0.5ex] 
 \hline\hline
 \MyIndent \textbf{Nr of reviews} & \textbf{57 369}    \\ 
 \MyIndent Male reviewer & 18 958  \\
 \MyIndent Female reviewer & 28 984   \\
 \MyIndent Unknown & 9 427   \\ \hline
 
 \MyIndent \textbf{Nr of different titles} & \textbf{14 647}   \\  
 \MyIndent Male author & 8 056  \\
 \MyIndent Female author & 6 591   \\ \hline
 
 \MyIndent \textbf{Nr of reviews by media type} &     \\ 
 \MyIndent Newspapers & 22 131  \\
 \MyIndent Blogs & 16 791   \\ 
 \MyIndent Online media & 10 635   \\ 
 \MyIndent Blog-like websites & 3 456   \\  
 \MyIndent Regional newspapers & 2 622   \\ 
 \MyIndent Weekly magazines & 1 566   \\ 
 \MyIndent Professional magazines & 168   \\ 
 [1ex] 
 \hline
 \end{tabular}
}{%
  \caption{An overview of the dataset presented in this paper. The category Online media includes (literary) sites that fall between online newspapers and personal blogs.}
\label{tab:overview}
}
\end{floatrow}
\end{figure}

The data set covers book reviews published in Danish media in the years 2010-2021. The data are retrieved from the online platform \href{https://bog.nu/}{bog.nu}'s API which collects book reviews published in Danish media. This includes reviews written in national newspapers, literature magazines, online media as well as in personal blogs. See table \ref{tab:overview} for a brief overview of the data set. 

\subsection{Grade Transformation and Estimation}

As different media use different grading scale, the grades on \texttt{bog.nu} are transformed to a 100-point grading scale. This approach, however, results in a sparse distribution of grades as the use of the original grading scales maps onto different intervals on the 100 point scale. Instead of this naive approach, we have used the original grade and applied a linear transformation to map all grades to a shared 6-point scale\footnote{It should be noted that most six-point scales have become the standard in many review outlets.}. Mapping from an $a$-$b$--point scale to a $1$-$6$--point scale: 

\begin{align}
 Y &= (B - A) \frac{(x - a)}{(b - a)} + A
  = (6 - 1) \frac{(x - a)}{(b - a)} + 1
  = \frac{5(x-a)}{b-a} +1
\end{align}

Figure \ref{fig:dist} shows the distributions of grades in Danish Newspapers transformed to a shared 6--point scale.

Some media do not provide a grade in a given review, but only a qualitative review. Bog.nu does however provide a quantification of the review, which is estimated by a human editor. For reviews written in Danish newspapers this estimation procedure is used in less than $25\%$ of the cases. Two important clarifications are needed: First, these estimates are made for both genders of both reviewers and authors. Secondly, to test the robustness of these estimates, the analysis below was performed both on the full data set and on the subset with original quantitative grades given in the reviews. We see that the same trends occur when excluding the reviews with estimated grades.

\subsection{Gender Retrieval}

The original data set from \texttt{bog.nu} does not contain gender for all authors and reviewers. We used a gendered name list to retrieve the missing gender variables. We are working with a binary understanding of gender, and we have used the API \href{https://genderize.io/}{genderize.io} that gives the probability of a name being either male or female, based on a data set of 250.000 names. We are aware of the problems with this method and how it rules out other gender identities \cite{harms_gender_exclusion}. However, a binary understanding of gender is still dominant in Denmark, and to investigate existing structures a binary gender variable is relevant.

\begin{figure}[h]
    \centering
    \includegraphics[width=8cm]{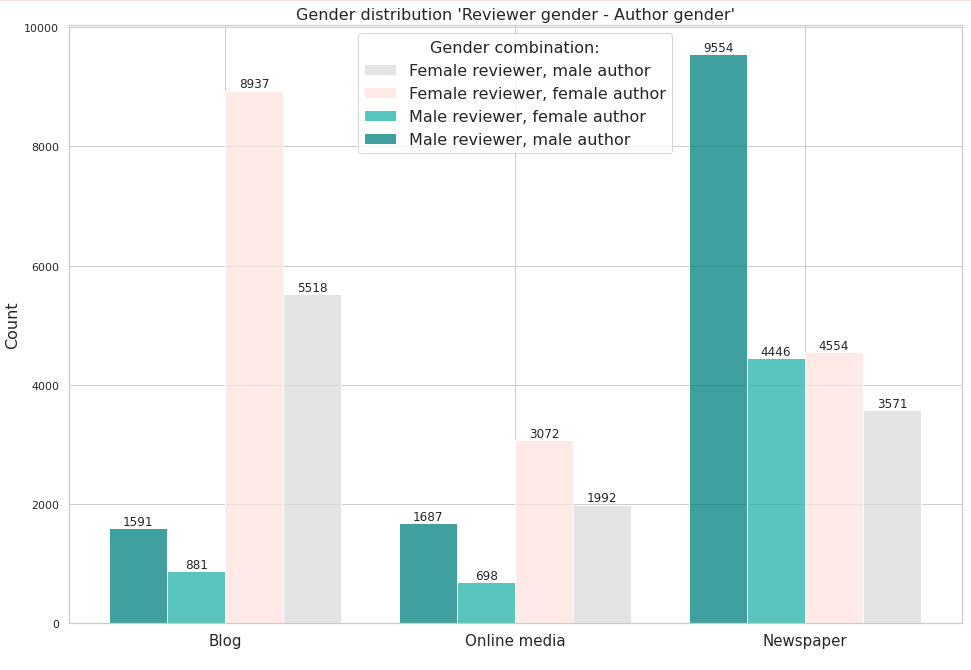} 
\caption{The histogram shows the distribution of reviewer- and author gender across the media types blog, online media, and newspapers.}
    \label{fig:grades}
\end{figure}

Looking at feature distributions in our data set, we see that both gender variables and grades differ across media types. As shown in Figure \ref{fig:grades}, we see a highly skewed gender distribution across media types: Male reviewers reviewing male authors are the dominant group in newspapers, whereas female reviewers reviewing female authors are the dominant group in the blogosphere. Furthermore, we have identified different `grading behaviours' in newspapers and in blogs. Hence, due to different distributions of gender as well as grades given across media types, we have in the rest of this study directed our attention towards newspapers.

\subsection{Model}

In order to estimate the effect the relative effect of author and reviewer gender of author on reviewer assigned grade (six-point scale), we solve the following linear model:

\begin{equation}
    y^{grade}_i = x_i \beta_{author} * x_i \beta_{reviewer} + \epsilon_i 
\end{equation}

Where $y_i$ is the grade of review $i$, $x_i$ is the predictor value (gender) of review $i$, $beta$ represents unknown parameters and $\epsilon$ is the error terms.

\section{Results}

\begin{figure}[ht]
    \centering
    \includegraphics[width=.49\textwidth]{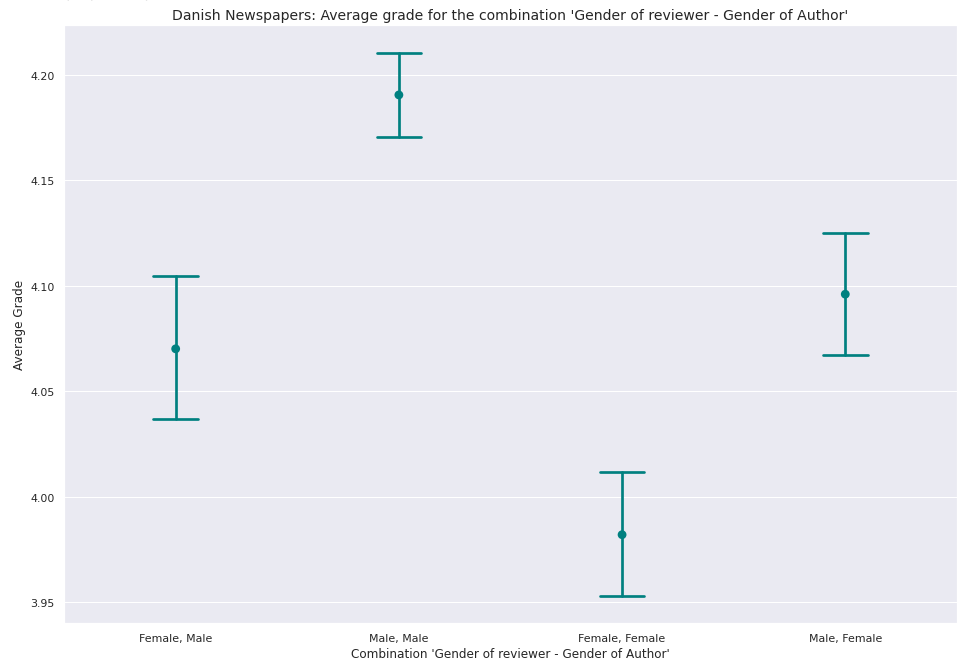}
    \includegraphics[width=.49\textwidth]{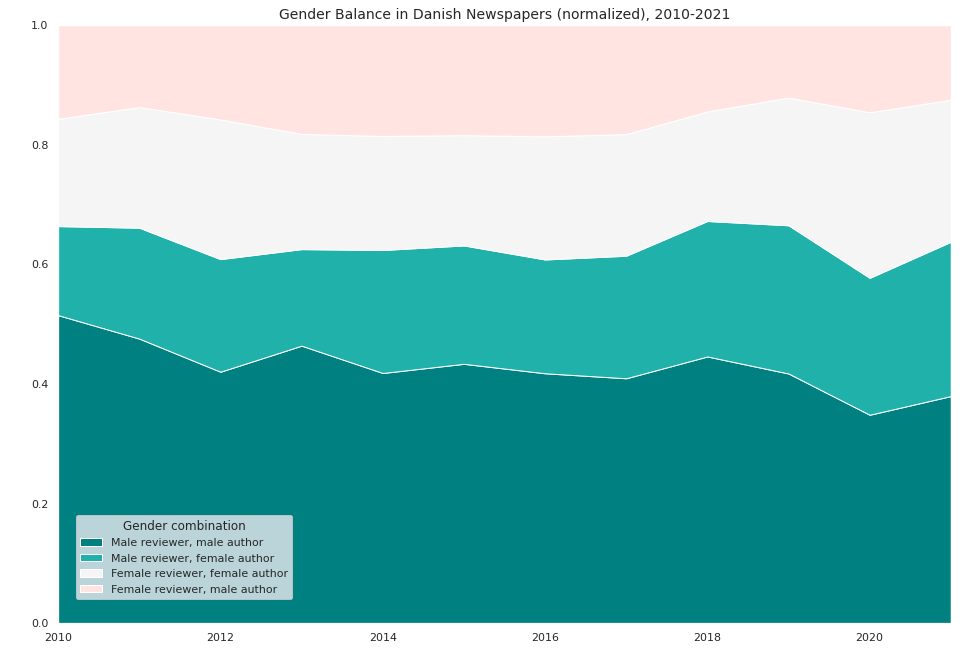} 
    \caption{The left-hand figure is a point plot showing the average grades in each of the four gender combinations. The lines indicate the standard deviations with a confidence interval of 95$\%$.
    The stacked areas plot on the right shows the percentage distribution over the years 2010-2021 of the four different gender combinations.}
    \label{fig:gender_dist}
\end{figure}

Solving for $y_i$ (grade of review) in the model above with ordinary least square (OLS), we get the following results, see Table \ref{tab:result}. The model and all contrasts are statistically significant ($p < .0001$) Conceptually, same gender (male * male, female * female) reviews spans the extreme values, while opposite gender (female * male, male * female) represent the middle of the distribution, see Fig. \ref{fig:gender_dist}. Female reviewers reviewing female authors account for the on average lowest grade. Male reviewers reviewing male authors results in the highest grade, with a 0.2 average grade increase. Opposite gender reviews is statistically speaking indistinguishable, but they differ on average 0.1 grade point from from same gender scoring.

\begin{table}[ht]
\begin{center}
\begin{tabular}{ l | ccccc }
Gender combination (reviewer * author)& Average grade & $SD$ & $ t $ & $p < |t| $ & $CI95\%$ \\
 \hline
Intercept {[}female * female{]} & 3.9820        & 0.015     & 267.502 & 0.0001              & {[}3.953, 4.011{]}    \\
female * male                  & 0.0881        & 0.022     & 3.924   & 0.0001              & {[}0.044, 0.132{]}    \\
male * female                  & 0.1140        & 0.021     & 5.382   & 0.0001              & {[}0.072, 0.156{]}    \\
male * male                    & 0.2084        & 0.018     & 11.523  & 0.0001              & {[}0.173, 0.244{]}   
\end{tabular}
\caption{Results for an OLS predicting grade based on the gender combination: reviewer and author.
The intercept is the gender combination female reviewer and female author. The $t$-test shows that all results are statistically significant. }
\label{tab:result}
\end{center}
\end{table}

Finally, as mentioned in section \ref{sec:methods} and shown in figure \ref{fig:grades}, men account for the majority of reviewers in the newspapers. Men actually dominate in number of reviewers (63$\%$ are men) and out of these, reviews are 69$\%$ reviews by male authors. Fig. \ref{fig:gender_dist} shows the development over the years 2010-2021. Here we see that the fraction of female authors being reviewed are slightly increasing, but the fraction of male reviewers is relatively stable through the years. 



\section{Discussion}

In line with the results in \cite{thelwall_2019, touileb-etal-2020-gender}, our results show that the gender of authors as well as of reviewers play a role in newspapers' literary reviews. In particular, the results above show that men review same gender more positive than opposite gender, and that women show the reverse pattern, that is, same gender is reviewed more negatively than opposite gender. A partial explanation of the behavior is that if males display a same gender preference and male reviewers make up the majority of newspaper reviewers, then the gender minority adapt to this preference and develop an same gender antagonism. At a more general level, the same gender preference of male in aesthetic judgement may reflect a cultural gender antagonism that follows a long historical trajectory. The female opposite gender preference is very likely to follow a the same cultural gender antagonism.  

There are caveats to this interpretation. First, it is possible that the gender bias is confounded with expertise bias, that is that specific literary language leads to higher literary appreciation. If women in general write more genre literature, then the observed difference may stem from a difference in complexity of linguistic features. In order to resolve this, we would need genre classification for all reviewed books. 
Second, although the average differences in grades are highly significant, the effect size is not considerable  (ex. 0.2 points on a six-point scale for same gender). This however begs the question, how large is a systematic differences supposed to be before it counts as a bias? We would argue that whenever we find systematic variation that co-insides with demographic variables, we are likely to see an indication of a relevant bias irrespective of the effect size. Conversely, if only a difference of a large magnitude (ex. two to three points on a six-point scale) were to count, then biases would only reflect common-sense propositions that most of us would share irrespective of their truth (ex. if women were on average reviewed two to three points lower, most of us would agree that they were worse writers).  

The last caveat points to an important issue, we are not arguing that this or that newspaper, or for that matter all newspapers, follow an explicit exclusionary strategy formulated by male reviewers and editors. Whenever we make a judgement there are essentially two sources of errors, bias and noise. While noise is randomly distributed and lack a systematic explanatory mechanism, biases are systematic and can be explained in terms of a mechanism. Demographic biases often originate in a systemic oppression of minority groups. For the specific review cases, the results are likely to mirror existing oppressive structures in the society such as those found in \cite{dane2020gender,macnell2015s, johnson2020dual, goldin2000orchestrating, booth2010employers, cole2004interaction}. We expect that majority groups in general will define norms and values that result in biased judgements irrespective of societal domain. Denmark is an excellent example of how more or less explicit biases persist in the face of societal countermeasures (ex. equal rights, parity policies) and in general high gender equality.


\section*{Acknowledgments}

This was was supported in part by the Velux Foundation for the Fabula-Net project and DeiC Interactive HPC with id DeiC-AU1-L-000011. The authors would like to thank \texttt{Bog.nu} for access to data.

\bibliographystyle{unsrt}  
\bibliography{references}

\end{document}